\documentclass[prd,twocolumn,showpacs,preprintnumbers,
amsmath,amssymb,superscriptaddress]{revtex4}
\usepackage{graphicx, epsfig, rotating, dcolumn, bm}
\usepackage{xspace}

\def\ss{\hat{s}}
\def\s1{\hat{s}_1}
\def\T1{\hat{t}_1}
\def\t2{\hat{t}_2}
\def\U1{\hat{u}_1}
\def\u2{\hat{u}_2}

\begin{document}
%%%%%%%%%%%%%%%%%%%%%%%%%%%%%%%%%%%%%%%%%%%%%%%%%%%%%%%%%%%%%%%%%%%%%%%%%%%%%%%

%\preprint{arXiv:yymm.nnnn}
\preprint{TIFR-TH/12-07}

\title{Azimuthal Angle Probe of Anomalous ${\bm{HWW}}$ Couplings at the LHeC}
%%%%%
\author{Sudhansu S. Biswal}
%\email{sudhansu.biswal@gmail.com}
\affiliation{College of Basic Sciences, Orissa University of Agriculture and
Technology, Bhubaneswar 751 003, India.}
\author{Rohini M. Godbole}
%\email{rohini@cts.iisc.ernet.in}
\affiliation{Center for High Energy Physics, Indian Institute of 
Science, Bangalore 560 012, India.}
\author{Bruce Mellado}
%\email{bmellado@mail.cern.ch}
\affiliation{Department of Physics, University of Wisconsin, 
Madison, WI 53706, USA. }
% \\
%{\em and}
\affiliation {University of the Witwatresrand, School of Physics,
Private Bag 3, Wits 2050, South Africa. }
\author{Sreerup Raychaudhuri}
%\email{sreerup@theory.tifr.res.in}
\affiliation{Department of Theoretical Physics, Tata Institute of Fundamental
Research, Mumbai 400~005, India.}

%------------------------------------------------------------------------------
\begin{abstract} 
A high energy $ep$ collider, such as the proposed LHeC, possesses the 
unique facility of permitting direct measurement of the $HWW$ coupling 
without contamination from the $HZZ$ coupling. At such a machine, the 
fusion of two $W$ bosons through the $HWW$ vertex would give rise to 
typical charged current (CC) events accompanied by a Higgs boson. We 
demonstrate that azimuthal angle correlations between the observable CC 
final states could then be a sensitive probe of the nature of the $HWW$ 
vertex and hence of the $CP$ properties of the Higgs boson.
\end{abstract} 
%------------------------------------------------------------------------------
\pacs{14.80.Cp, 13.60.-r, 11.30.Er} 
\maketitle

The Higgs boson has long been sought for as the cornerstone to the 
entire mechanism of electroweak symmetry-breaking \cite{EWSB} in the 
Standard Model (SM) \cite{GSW}. The hunt has been long and frustrating, 
but since the announcement of the latest search results by the 
experimental collaborations \cite{Higgs_search}, we now know that a new 
boson has been found with a mass around 126\, GeV and that this boson 
resembles the Higgs boson of the SM. By the end of the current year we 
may have enough data to identify this particle as a Higgs boson $H^0$ 
with couplings proportional to mass --- which in turn, will provide very 
convincing evidence that the electroweak symmetry is indeed 
spontaneously broken through a scalar doublet $\Phi$ acquiring nonzero 
vacuum expectation value. However, mere identification as a Higgs boson 
is not enough, for it will leave open a host of other questions, such as 
whether this scalar is elementary or composite, $CP$-conserving or 
$CP$-violating, and so on. Of course, the minimal SM has only one 
physical scalar $H^0$, with $J^{PC} = 0^{++}$, but this, like so much 
else in the SM, is essentially an ad hoc assumption made with a view 
towards economy of fields and interactions, rather than the product of 
any deeper understanding of the underlying physics. It will, therefore, 
be necessary to test the spin and $CP$ properties of the new boson 
experimentally, before we can truly identify it with the Higgs boson of 
the SM.

The all-important question of how the symmetry-breaking is transmitted 
from the scalar sector to the gauge sector is answered in the SM by 
having gauge boson-scalar couplings arising from the assignment of 
non-trivial gauge quantum numbers to the scalar fields in the theory. As 
a result, the couplings of the $H^0$ to the heavy electroweak gauge 
bosons $W^\pm$ and $Z^0$ are precisely formulated in the SM, and come 
out as
\cite{GSW}
\begin{equation}
{\cal L}_{\rm int} = - g M_W \left(W_\mu W^\mu + \frac{1}{2\cos\theta_W} 
Z_\mu Z^\mu \right) H
\label{eqn:HWWSM}
\end{equation}
Since $g$, $M_W$ and $\theta_W$ are all accurately measured, this vertex 
is fully determined in the SM. However, if we wish to confirm that the 
SM mechanism for breaking electroweak symmetry is the correct one, we 
would require an independent measurement of these vertices. This is 
easier said than done, though, because ($a$) one will require to produce 
a substantial number of Higgs bosons through these electroweak vertices, 
which would require accumulation of considerable statistics before a 
precision result can be claimed, and more importantly, ($b$) because 
these vertices are sensitive to the presence of new physics beyond the 
SM, with corrections occurring mostly at the one-loop level. If we 
parametrise the $H(k)-W^+_\mu(p)-W^-_\nu(q)$ vertex in the general form
\begin{equation}
i\Gamma^{\mu\nu}(p,q)\ \epsilon_\mu(p)\ \epsilon^\ast_\nu(q) \ ,
\end{equation} 
any deviations from the simple SM formula $\Gamma_{\rm 
(SM)}^{\mu\nu}(p,q) = - gM_W\, g^{\mu\nu}$ in Eqn.~(\ref{eqn:HWWSM}) -- 
at a level incompatible with SM radiative corrections -- would 
immediately indicate the presence of new physics beyond the SM (BSM). 
Following Ref.~\cite{PleRaiZep}, we can parametrise these deviations 
using two dimension-5 operators
\begin{equation}
\Gamma^{\rm BSM}_{\mu\nu}(p,q) = \frac{g}{M_W}\left[ 
\lambda \left(p \cdot q\, g_{\mu\nu} - p_\nu q_\mu  \right)
+ i\, \lambda^\prime\ \epsilon_{\mu\nu\rho\sigma}p^\rho q^\sigma
\right]
\label{eqn:BSMvertex}
\end{equation}
where $\lambda$ and $\lambda^\prime$ are, respectively, effective 
coupling strengths for the anomalous $CP$-conserving and the 
$CP$-violating operators. One can make \cite{PleRaiZep} a similar 
parametrisation for the $H(k)-Z_\mu(p)-Z_\nu(q)$ vertex, with another 
pair of unknown couplings $\widetilde{\lambda}$ and 
$\widetilde{\lambda}^\prime$ and the replacement $M_W \to M_Z$. We can 
even have a $H\gamma\gamma$ vertex with yet another pair of unknown 
couplings~\cite{Hgamma}. This last will vanish in the SM at tree level, 
but it certainly appears at the one-loop level, where it is well known 
to provide one of the cleanest channels \cite{Higgs_search} to search 
for the $H^0$.

%==================================
\begin{figure}[!h]
\begin{center}
\includegraphics[width = 0.35\textwidth]{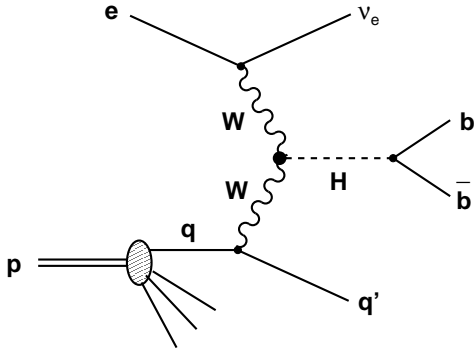} 
\end{center}
\caption[]{Higgs boson production at an $ep$ collider through 
$WW$ fusion and the $HWW$ vertex.}
\label{fig:diagram}
\end{figure}
%==================================

The above parametrisation of anomalous $HWW$ and similar couplings 
illustrates the important point that the $CP$ properties of the Higgs 
boson are rather difficult to measure directly, but will be known if we 
can determine the couplings $\lambda$ and $\lambda'$ to any degree of 
certainty. A survey of the literature throws up several suggestions 
\cite{HiggsCP, Zeppenfeld, Biswal} on how this can be done at colliders, 
mostly using angular correlations between the final states. An 
additional complication arises, however, as all the observables studied 
so far in the context of hadronic colliders 
\cite{CPNSH,HiggsCP,Zeppenfeld} as well as the electron positron 
colliders \cite{CPNSH, Biswal}, are dependent on {\it more than one} of 
these couplings. Thus, even if a deviation from the SM prediction is 
observed, it will be difficult to disentangle the responsible vertex in 
such studies \cite{Zeppenfeld, Biswal,Ruwiedel}. As pointed out in 
Refs.~\cite{Rindani,Gunion} a study of $e^+ e^- \rightarrow t \bar t 
H^0$ production offers the possibility of a clear and unambiguous 
determination of the CP properties of the $H^0$; however, at the LHC 
this process may be accessible only in the high energy and luminosity 
phase. However, it is interesting to note that the production of a Higgs 
boson in the WW fusion process in the charged current reactions $e + p 
\rightarrow \nu H^0 X$ \cite{LHeC,HanMel} or $\nu + p \rightarrow e H^0 
X$ \cite{Godbole} arise only from a single Feynman diagram involving the 
$HWW$ vertex as shown in the Figure~\ref{fig:diagram} for $e+ p 
\rightarrow \nu_e + X + H (b \bar b)$. These modified charged current 
(CC) processes not only provide the best way to observe the $H \to 
b\bar{b}$ decay, but also render the measurement of the $HWW$ vertex 
free from possible contamination by contributions from $HZZ$ or 
$H\gamma\gamma$ vertices. Moreover, the $ep$ collision has an additional 
advantage over the LHC in that the initial states would be asymmetric. 
Thus, we can disentangle backward scattering from forward scattering and 
study these separately, which is not possible at the LHC. In this 
letter, therefore, we focus on the measurement of the $HWW$ vertex in 
such CC events at the high-energy high-luminosity $ep$ collider 
envisaged in the LHeC proposal \cite{LHeC}, where a high energy ($\sim 
50 - 150$~GeV) beam of electrons would be made to collide with the 
multi-TeV beams from the LHC. Such a machine will have a centre-of-mass 
energy as high as $1 - 1.5$~TeV and can therefore produce $H^0$ events 
copiously \cite{LHeC,HanMel}.

A glance at Figure~\ref{fig:diagram} will show that the final state has 
missing transverse energy (MET) and three jets $J_1$, $J_2$ and $J_3$, 
of which two (say $J_2$ and $J_3$) can be tagged as $b$-jets.  At the 
parton level, the squared and spin-summed-averaged matrix element for 
the process
$$
e^-(k_1) + q(k_2) \longrightarrow \nu_e(p_1) + q'(p_2) + H(p_3)
$$
can now be worked out to be
\begin{widetext}
\begin{eqnarray}
&\overline{\left|{\cal M}\right|^2}&  
= \left( \frac{4\pi^3\alpha^3}{\sin^6\theta_W} \right) 
\frac{1}{M_W^2(\T1 - M_W^2)^2\,(\u2 - M_W^2)^2}  \times\bigg[ 4 M_W^4\, \ss \s1   \nonumber \\
&+ & \hskip -10pt \lambda^2   \left\{ \T1 \u2 ( \ss^2 + \s1^2 + \T1 \u2- 2 \t2 \U1) + (\ss \s1 - \t2 \U1)^2 \right\}  
+  2 \lambda M_W^2 ( \ss + \s1) (\ss \s1 + \T1 \u2 - \t2 \U1)  \nonumber \\
& + & \hskip -10pt \lambda'^2 \left\{ \T1 \u2 (\ss^2 + \s1^2 -\T1 \u2 + 2 \t2 \U1) - (\ss \s1 - \t2 \U1)^2  \right\}  
- 2 \lambda' M_W^2 ( \ss - \s1) (\ss \s1 + \T1 \u2 - \t2 \U1) \nonumber \\
& + & \hskip -10pt 2 \lambda\lambda' \T1 \u2 (\s1^2 - \ss^2)  \bigg] 
\label{eqn:matrixelement}
\end{eqnarray}
\end{widetext}
where the invariant variables are defined by $\ss = (k_1 + k_2)^2$, $\T1 
= (k_1 - p_1)^2$, $\U1 = (k_1 - p_2)^2$, $\s1 = (p_1 + p_2)^2$, $\t2 = 
(k_2 - p_1)^2$ and $\u2 = (k_2 - p_2)^2$. The first term inside the 
square brackets is the SM contribution and is, of course, just the beta 
decay matrix element. The other terms include direct and interference 
BSM contributions of both $CP$-conserving and $CP$-violating types and 
even a crossed term between the two types of BSM contributions.

\begin{figure}[!h]
\begin{center}
\includegraphics[width=0.35\textwidth]{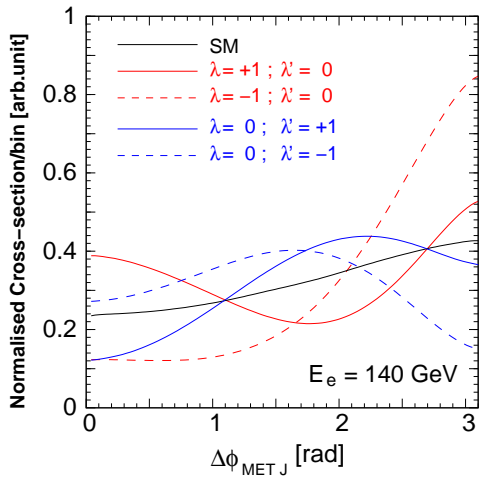} 
\end{center}
\caption[]{Azimuthal angle distributions in the SM and with 
anomalous $HWW$ couplings.}
\label{fig:theory}
\end{figure}

The expression in Eqn.~(\ref{eqn:matrixelement}), though exact, is not 
very transparent. It can be shown \cite{PleRaiZep}, however, that in the 
limit when there is practically no energy transfer to the $W$ bosons and 
the final states are very forward, the $CP$-conserving ($CP$-violating) 
coupling $\lambda$ ($\lambda^\prime$) contributes to the matrix element 
for this process a term of the form
\begin{equation}
{\cal M}_\lambda \ \propto \ +\lambda \, \vec{p}_{T1}.\vec{p}_{T2}
\qquad\qquad
{\cal M}'_\lambda \ \propto \  -\lambda^\prime \, \vec{p}_{T1}.\vec{p}_{T2} \ ,
\label{eqn:approx}
\end{equation}
where $\vec{p}_{T1}$ is the vector of the missing transverse energy. 
These terms ${\cal M}_\lambda$ and ${\cal M}'_\lambda$ both go through a 
zero when the azimuthal angle $\Delta \varphi_{\rm MET-J}$ between the 
non-$b$ jet $J_1$ (arising from the parton $q'$) and the missing 
transverse energy is $\pi/2$ or $3\pi/2$. When ${\cal M}_\lambda$ and 
${\cal M}'_\lambda$ are added to the relatively flat (in $\Delta 
\varphi_{\rm MET-J}$) SM background, one predicts a curve with a peak 
(dip) around $\Delta \varphi_{\rm MET-J} \approx 0 (\pi)$ for the 
$\lambda$ operator and the opposite behaviour for the $\lambda^\prime$ 
operator, when the signs of $\lambda, \lambda^\prime$ are positive and 
vice versa when they are negative. The exact behaviour is illustrated in 
Figure~\ref{fig:theory}, which was generated for the case of a 140~GeV 
electron colliding with a 6.5~TeV proton and setting the Higgs boson 
mass to 125~GeV. Since the approximations which reduce 
Eqn.~(\ref{eqn:matrixelement}) to Eqn.~(\ref{eqn:approx}) are somewhat 
too drastic, these curves show the expected qualitative behaviour but 
the peaks (dips) are somewhat displaced from the values quoted above.

%==================================
%================================

In generating these `theoretical' distributions, no kinematic cuts were 
applied. The choices of $\lambda, \lambda' = 0, \pm 1$ in 
Figure~\ref{fig:theory} are completely ad hoc -- in a specific BSM model 
the actual value can vary considerably -- but they serve the purposes of 
illustration well.  Of course, the precise value of $\lambda$ (or 
$\lambda^\prime$) is crucial to any actual study -- in the limit 
$\lambda \to 0$ (or $\lambda^\prime \to 0$) we would naturally get 
distributions which are practically indistinguishable from the SM 
prediction. In our subsequent analysis, we shall see how we can 
constrain the values of $\lambda, \lambda'$ is a model-independent way. 
We find it convenient to study the cases of $CP$ conserving anomalous 
couplings and $CP$-violating anomalous couplings separately, for the 
$CP$-conserving $\lambda$ term will be generated even in the SM at 
one-loop level, whereas the $CP$-violating $\lambda^\prime$ will arise 
at this order only if there is new BSM physics. Thus, in 
Figure~\ref{fig:theory}, we consider $\lambda \neq 0$ when 
$\lambda^\prime = 0$ and vice versa.

In this, and the subsequent numerical analysis, we are careful to use 
the exact formulae in Eqn.~(\ref{eqn:matrixelement}), convoluted with 
parton density functions from the CTEQ6L set \cite{CTEQ} as well as the 
MSTW-2008 set \cite{MSTW}. PDF errors were estimated by running over all 
the available CTEQ6L and MSTW LO data sets. We found that Hessian errors 
and differences in fitting techniques between CTEQ and MSTW PDFs do lead 
to fairly significant overall changes in the overall cross-section, but 
when it comes to the normalised distribution in azimuthal angle of 
Figure~\ref{fig:theory}, the differences turn out to be so small that 
they can practically be absorbed in the thickness of the lines shown on 
Figure~\ref{fig:theory}. We do not, therefore, include PDF uncertainties 
in our error analysis. It is also worth noting that if we vary the Higgs 
boson mass between $120 - 130$~GeV, the production cross-section changes 
somewhat, but again this hardly affects the normalised distributions 
shown in Figure~\ref{fig:theory}.

In order to go beyond the simple-minded parton-level study, however, it 
is necessary to apply kinematic cuts and simulate the fragmentation of 
the partons to jets, before a realistic estimate of the sensitivity of 
this process to $\lambda$ and $\lambda'$ can be estimated. These effects 
tend to distort the characteristic curves shown in 
Figure~\ref{fig:theory} -- but not enough to disrupt their qualitative 
differences. Instead of making a detailed simulation of the 
fragmentation processes, however, we have smeared the partonic energies 
with the hadronic energy relative resolution $\sigma_E/E = 
\sqrt{\alpha^2/E + \beta^2}$ where $\alpha = 0.6$~GeV$^{1/2}$ and $\beta 
= 0.03$. This leads to a resolution of about 7\% on the invariant mass 
of the Higgs boson if we do not smear the angular distribution of the 
jets. Once this is done, we have made a detailed simulation based on the 
exact kinematic criteria and efficiencies adopted in Ref.~\cite{HanMel}, 
which studies the same process from the point of view of determining $Hb 
\bar b$ coupling for a SM Higgs boson. These criteria may be summarised 
as follows:
\begin{enumerate}
\item It is required that $MET>25\,$GeV.
\item Presence of two $b$-partons with $p_T^b > 30$~GeV and $|\eta_b| < 
2.5$. The invariant mass of these $b$-partons must lie within 10~GeV of 
the Higgs boson mass.
\item Of the remaining partons, the leading one must have $p_T > 30$~GeV 
and $1 < \eta < 5$. This will be called the forward tagging parton.
\item We require $\Delta \varphi_{\rm MET-J} > 0.2\,$rad for all 
the jets (J).
\item A veto on leptons ($\ell = e, \mu, \tau$) with $p_T^\ell
> 10$~GeV and $|\eta_\ell| < 2.5$ is required.
\item The invariant mass of the Higgs boson candidate and the forward tagging 
jet must be greater than 250~GeV.
\item $b$-tagging efficiency: $\varepsilon_b = 0.6$ for $|\eta_b| < 
2.5$. The mis-tagging factor for $c$ and light quark jets is taken as 
0.1 and 0.01 respectively. 
\end{enumerate}

Taking all these criteria, the azimuthal distribution has been simulated 
in 10 bins, each of width $\pi/5$, and the signal for each value of 
$\lambda$ ($\lambda'$) and SM backgrounds have been calculated in each 
bin using the same formulae used to create Figure~\ref{fig:theory}.  
Assuming statistical errors dependent on the integrated luminosity, $L$, 
we then determine the sensitivity, for a given $L$, of the experiment to 
$\lambda, \lambda^\prime$ by making a log-likelihood analysis. The 
background estimation has been taken from the studies described in 
Ref.~\cite{Klein}. It may be noted that these criteria are optimised for 
a Higgs mass of 120~GeV, as in in Ref.~\cite{HanMel}, and could change 
marginally for the favoured range set by the experimental 
collaborations~\cite{Higgs_search}. However, such changes hardly matter 
for the present analysis.

\begin{figure*}[ht!]
\begin{center}
\includegraphics[height=5.6cm,width=12.5cm]{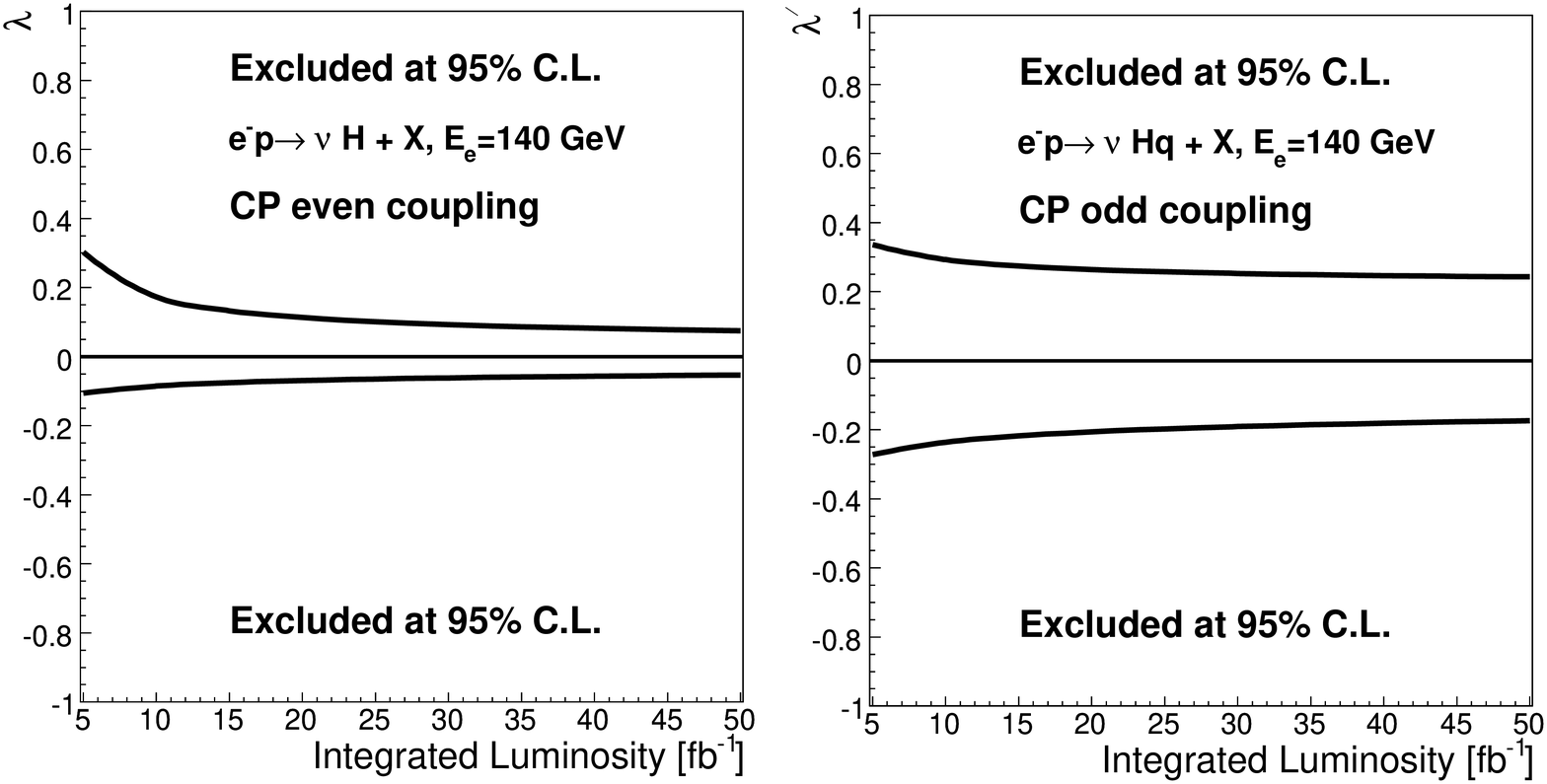} 
\end{center}
\caption[]{Exclusion plots obtainable by a study of the azimuthal angle 
distributions at the LHeC for the $CP$-even coupling $\lambda$ and the 
$CP$-odd coupling $\lambda^\prime$.}
\label{fig:simul}
\end{figure*}

Our results are exhibited in Figure~\ref{fig:simul}, where we present 
95\% exclusion plots for the anomalous couplings as a function of $L$. 
The left panel shows the exclusion plot for $\lambda$, while the right 
shows the exclusion plot for $\lambda'$. It is clear from this figure 
that by the time the LHeC has collected 10~fb$^{-1}$ of data, we will be 
able to discover anomalous couplings down to the level of 0.3 or lower, 
or else to exclude such couplings and establish to that extent that the 
$HWW$ vertex indeed resembles the SM vertex.  We note that the process 
in question is somewhat more sensitive to the $CP$-even coupling, as 
evidenced by the narrower inaccessible region indicated on the left 
panel.

%==================================
%==================================

It is interesting to ask what happens if the energy of the electron beam 
is different from 140~GeV, as assumed in the previous discussion. The 
azimuthal angle distributions shown in Figure~\ref{fig:theory} hardly 
change as the electron beam energy $E_e$ is changed through 50~GeV to 
200~GeV. The acceptance of the CC Higgs boson signal has been evaluated 
in~\cite{HanMel}. If $E_e$ is decreased while keeping the energy of the 
proton beam constant, the acceptance decreases minimally so long as 
$E_e$ is above 100~GeV, but begins to decrease significantly for $E_e$ 
less than 100~GeV. The acceptance of the Higgs boson signal for $E_e = 
50$~GeV is, in fact, diminished by 25\% with respect to that of $E_e = 
100$~GeV. Most of this acceptance loss stems from the requirement of two 
$b$-jets. Part of the acceptance can be recovered by allowing for 
tracking and calorimeter coverage to increase in the forward direction.

In summary, the LHeC is the only machine where one can measure the $HWW$ 
coupling directly without making any prior assumptions about new BSM 
physics. We have shown that the azimuthal angle $\Delta \varphi_{\rm 
MET-J}$ in CC events accompanied by a $H$ boson at the LHeC is a 
powerful and unambiguous probe of anomalous $HWW$ couplings, both of the 
$CP$-conserving and and the $CP$-violating type, and is robust against 
uncertainties in the exact Higgs boson mass and the PDF errors.  We 
conclude that an integrated luminosity of around 10~fb$^{-1}$ would 
suffice to probe reasonably small values of these couplings.

This work was supported in part by the DOE Grant No. DE- FG0295-ER40896. 
RG wishes to thank the Department of Science and Technology, Government 
of India, for support under grant no. SR/S2/JCB-64/2007.

\end{document}